\batchmode
\makeatletter
\def\input@path{{D:/arxiv-2101.02426v2/}}
\makeatother
\documentclass[english,aps,preprint]{revtex4-1}
\usepackage[T1]{fontenc}
\usepackage{array}
\usepackage{verbatim}
\usepackage{longtable}
\usepackage{booktabs}
\usepackage{amsmath}
\usepackage{amssymb}
\usepackage{stmaryrd}

\makeatletter

\providecommand{\tabularnewline}{\\}

\makeatother

\usepackage{babel}
\begin{document}
\preprint{USTC-ICTS/PCFT-21-03}
\title{Class of Bell-Clauser-Horne inequalities for testing quantum nonlocality}
\author{Chen Qian}
\email{qianch18@mail.ustc.edu.cn}

\affiliation{Interdisciplinary Center for Theoretical Study and Department of Modern
Physics, University of Science and Technology of China, Hefei, Anhui
230026, China}
\affiliation{Peng Huanwu Center for Fundamental Theory, Hefei, Anhui 230026, China}
\author{Yang-Guang Yang}
\email{mathmuse@ustc.edu.cn}

\affiliation{Interdisciplinary Center for Theoretical Study and Department of Modern
Physics, University of Science and Technology of China, Hefei, Anhui
230026, China}
\affiliation{Peng Huanwu Center for Fundamental Theory, Hefei, Anhui 230026, China}
\author{Cong-Feng Qiao}
\email{qiaocf@ucas.ac.cn}

\affiliation{School of Physical Sciences, University of Chinese Academy of Sciences,
Beijing 100049, China}
\affiliation{Key Laboratory of Vacuum Physics, Chinese Academy of Sciences, Beijing
100049, China}
\author{Qun Wang}
\email{qunwang@ustc.edu.cn}

\affiliation{Interdisciplinary Center for Theoretical Study and Department of Modern
Physics, University of Science and Technology of China, Hefei, Anhui
230026, China}
\affiliation{Peng Huanwu Center for Fundamental Theory, Hefei, Anhui 230026, China}
\begin{abstract}
Quantum nonlocality, one of the most important features of quantum
mechanics, is normally connected in experiments with the violation
of Bell-Clauser-Horne (Bell-CH) inequalities. We propose effective
methods for the rearrangement and linear inequality to prove a large
variety of Bell-CH inequalities. We also derive a set of Bell-CH inequalities
by using these methods which can be violated in some quantum entangled
states.
\end{abstract}
\maketitle

\section{introduction}

\label{sec:introduction}Quantum nonlocality is one of the most striking
aspects of quantum mechanics without a classical analog in reality.
It can be characterized as correlated outcomes when we measure two
or more entangled quantum systems, even if these systems are spatially
separated. Quantum nonlocality originates from a contradiction between
local realism and the completeness of quantum mechanics pointed out
by Einstein, Podolsky and Rosen in 1935 \citep{Einstein1935}, which
was called the ``EPR paradox'' and led to a great challenge of the
concept of ``locality'' taken for granted by most physicists. To
test such a contradiction in a real physical system, Bell, Clauser,
and other researchers formulated observable inequalities with correlation
functions of measurement outcomes for entangled systems \citep{Bell1964,Clauser1969,Bell1971}.
These inequalities were later called Bell-Clauser-Horne-Shimony-Holt
(Bell-CHSH) inequalities. They proved that the quantum correlation
cannot be fully described by any local hidden-variable theory. Later
on, Clauser and Horne proposed the CH inequality in terms of probabilities
\citep{Clauser1974}, followed by many CH-like inequalities (we call
them all Bell-CH inequalities in this paper). These inequalities have
weaker auxiliary assumptions than Bell-CHSH inequalities in experimental
considerations.

Violation of Bell-CH inequalities has been widely verified by many
experiments in favor of the nonlocal feature of quantum mechanics
\citep{Aspect2002,Ding2007,Li2010,Qian2020}. However, observing the
violation of these inequalities is not feasible in every entangled
system \citep{Hiesmayr2015}. One needs to generalize these inequalities
in order to test them in experiments. The first attempt to find a
systematic way of generalization was made by Froissart \citep{Froissart1981},
and later attempt was made by Pitowsky \citep{Pitowsky1986,Pitowsky1989,Pitowsky1991},
who introduced correlation polytopes to find Bell-CH inequalities
inspired by Boole's method on probabilistic inequalities \citep{Pitowsky1994}.
After a series of further works, such algorithms were built to find
a set of Bell-CH inequalities based on the solution of convex problems
\citep{Froissart1981,Collins2004,Avis2005,Avis2006,Ito2006,Bancal2010,Pironio2014}.
A large variety of Bell-CH inequalities have been found \citep{Froissart1981,Sliwa2003,Collins2004,Ito2006,Brunner2008,Gisin2009,Pal2009,Bancal2010,Quintino2014,Cope2018,Oudot2019,Cruzeiro2019}.
In Refs. \citep{Cope2018,Oudot2019,Cruzeiro2019}, the authors obtained
a complete list of Bell-CH inequalities in the case of binary settings
up to four measurements on each setting. The main goal of this paper
is to prove these inequalities with analytical techniques and construct
a class of Bell-CH inequalities. Our inequalities can be used as a
substitute for the original ones for testing quantum nonlocality under
some circumstances.

We consider a system consisting of two classical or quantum subsystems
$x$ and $y$. Alice chooses $m$ measurement settings for a quantity
that has $k$ values on $x$, and Bob chooses $n$ measurement settings
for a quantity that has $l$ values on $y$. We denote this case as
$mnkl$. The system can be measured many times with different selected
settings. We define $P_{x_{i}}$ as the probability that Alice obtains
a certain value $a$ for measurement setting $i$ on $x$, $P_{y_{j}}$
as the probability that Bob obtains a certain value $b$ for measurement
setting $j$ on $y$, and $P_{x_{i}y_{j}}$ as the probability that
Alice obtains $a$ for $i$ and Bob obtains $b$ for $j$ at the same
time. For a classical system labeled $mn22$, we have the so-called
Bell-CH inequalities of binary settings which hold for certain integer
coefficients $C_{x_{i}}$, $C_{y_{j}}$, and $C_{x_{i}y_{j}}$,
\begin{equation}
\sum_{i=1}^{m}C_{x_{i}}P_{x_{i}}+\sum_{j=1}^{n}C_{y_{j}}P_{y_{j}}+\sum_{i=1}^{m}\sum_{j=1}^{n}C_{x_{i}y_{j}}P_{x_{i}y_{j}}\leqslant0.\label{eq:Bell-CH-ieq}
\end{equation}
In our formulation, we define the corresponding Bell-CH-like inequalities
in algebraic forms (we call them algebraic Bell-CH inequalities for
short) as
\begin{equation}
B\sum_{i=1}^{m}C_{x_{i}}x_{i}+A\sum_{j=1}^{n}C_{y_{j}}y_{j}+\sum_{i=1}^{m}\sum_{j=1}^{n}C_{x_{i}y_{j}}x_{i}y_{j}\leqslant0,\label{eq:Bell-CH-ieq-c}
\end{equation}
with $A$ and $B$ being the upper bounds of two sets of positive
variables $x_{i}$ and $y_{i}$, respectively. The Bell-CH inequalities
can be derived from the algebraic ones in local hidden-variable theory.
Among these inequalities, there are independent ones which we label
$Imn22$ for Bell-CH inequalities and \emph{$I_{mn22}$} for algebraic
ones.

This paper is organized as follows. In this section, we have provided
an introduction and the motivation of this work with definitions of
Bell-CH inequalities and their algebraic forms. In Sec. \ref{sec:algebraic-Bell-CH},
we derive algebraic Bell-CH inequalities with two methods; one is
from the rearrangement inequality, and the other is from the linear
inequality. In Sec. \ref{sec:Bell-CH}, we derive Bell-CH inequalities
and obtain a different class of inequalities. In Sec. \ref{sec:testing},
we test our inequalities with special quantum entangled states. In
Sec. \ref{sec:summary} we summarize this work and draw conclusions.

\section{Algebraic Bell-CH inequalities for binary settings}

\label{sec:algebraic-Bell-CH}

\subsection{Method of rearrangement inequality}

\label{subsec:method-of-rearrangement}The rearrangement inequality
is a well-known inequality in classical mathematics, from which one
can prove many famous inequalities, such as the arithmetic-mean–geometric-mean
inequality, Cauchy inequality, etc. In Appendix \ref{sec:rearrange}
we give a very brief introduction of the rearrangement inequality;
readers can see Ref. \citep{Hardy1988} for a review of this topic.

In this section, we derive two low-order algebraic Bell-CH inequalities
by applying the rearrangement inequality: algebraic Bell-CH inequalities
$I_{2222}$ and $I_{3322}$. In order to prove these inequalities,
we use the maximum and minimum values of variables to enlarge the
upper bound of polynomials in intermediate steps. Then we can rearrange
the orders of variables in these polynomials and apply the rearrangement
inequality to determine their signs. In this way, we find a profound
connection between these inequalities and the classical rearrangement
inequality.

\subsubsection{$I_{2222}$}

We define two sets of $m$ and $n$ positive real numbers
\begin{eqnarray}
a & = & \{x_{i},i=1,\cdots,m\},\nonumber \\
b & = & \{y_{j},j=1,\cdots,n\},\label{eq:s-eq}
\end{eqnarray}
with upper bounds $A$ and $B$, respectively,
\begin{align}
0 & \leqslant x_{i}\leqslant A,\ 0\leqslant y_{j}\leqslant B.\label{eq:bc}
\end{align}
The maximum and minimum values of the numbers in set $a$ are denoted
$x_{+}(1,\ldots,m)$ and $x_{-}(1,\ldots,m)$, respectively, and the
maximum and minimum values of the numbers in set $b$ are denoted
$y_{+}(1,\ldots,n)$ and $y_{-}(1,\ldots,n)$, respectively.

We define $I_{2}$ as a polynomial of the following form:
\begin{equation}
I_{2}=x_{1}y_{1}+x_{1}y_{2}+x_{2}y_{1}-x_{2}y_{2}-x_{1}B-Ay_{1},\label{eq:I2}
\end{equation}
which involves elements of set $a$ and $b$ in (\ref{eq:s-eq}) with
$m=n=2$. We now prove the $I_{2222}$ inequality, $I_{2}\leqslant0$.
Using boundary conditions
\begin{align}
0\leqslant x_{-}(1,2)\leqslant x_{1},x_{2} & \leqslant x_{+}(1,2)\leqslant A,\nonumber \\
0\leqslant y_{-}(1,2)\leqslant y_{1},y_{2} & \leqslant y_{+}(1,2)\leqslant B
\end{align}
in (\ref{eq:bc}), we obtain
\begin{eqnarray}
I_{2} & \leqslant & x_{1}(y_{1}+y_{2})+x_{2}(y_{1}-y_{2})-x_{1}y_{+}(1,2)-x_{+}(1,2)y_{1}\nonumber \\
 & \leqslant & x_{1}(y_{1}+y_{2})+x_{2}(y_{1}-y_{2})-x_{1}y_{+}(1,2)-x_{+}(1,2)y_{1}\nonumber \\
 &  & -\left[x_{1}-x_{+}(1,2)\right]y_{-}(1,2)-x_{-}(1,2)\left[y_{1}-y_{+}(1,2)\right],\label{eq:I2-e-1}
\end{eqnarray}
where we have replaced $A$ and $B$ with $x_{+}(1,2)$ and $y_{+}(1,2)$
in Eq. (\ref{eq:I2}), respectively, to obtain the first line and
added two positive quantities in the second inequality.

Since there are only two variables in each set, we always have
\begin{align}
x_{+}(1,2)+x_{-}(1,2) & =x_{1}+x_{2},\nonumber \\
y_{+}(1,2)+y_{-}(1,2) & =y_{1}+y_{2}.
\end{align}
Using the above identity, inequality (\ref{eq:I2-e-1}) becomes
\begin{eqnarray}
I_{2} & \leqslant & x_{1}(y_{1}+y_{2})+x_{2}(y_{1}-y_{2})-x_{1}(y_{1}+y_{2})-(x_{1}+x_{2})y_{1}\nonumber \\
 &  & +x_{+}(1,2)y_{-}(1,2)+x_{-}(1,2)y_{+}(1,2)\nonumber \\
 & = & -(x_{1}y_{1}+x_{2}y_{2})+x_{+}(1,2)y_{-}(1,2)+x_{-}(1,2)y_{+}(1,2)\nonumber \\
 & \equiv & I_{2}^{(0)}.\label{eq:I2-e-2}
\end{eqnarray}
We see in (\ref{eq:I2-e-2}) that $I_{2}^{(0)}$ is the difference
between the reversed sum and the unordered sum for sets $a$ and $b$
in (\ref{eq:s-eq}) with $m=n=2$. Using the rearrangement inequality,
we obtain
\begin{equation}
I_{2}\leqslant I_{2}^{(0)}\leqslant0.\label{eq:I2-e-r}
\end{equation}
This concludes the proof of the $I_{2222}$ inequality by applying
the rearrangement inequality.

We can rewrite $I_{2}^{(0)}$ with the help of the Heaviside step
function,
\begin{eqnarray}
\theta(x) & = & \begin{cases}
\begin{array}{c}
1\\
\frac{1}{2}\\
0
\end{array} & \begin{array}{c}
x>0,\\
x=0,\\
x<0.
\end{array}\end{cases}\label{eq:theta}
\end{eqnarray}
The maximum and minimum values can be put into the following form:
\begin{align}
x_{+}(1,2) & =x_{1}\theta(x_{1}-x_{2})+x_{2}\theta(-x_{1}+x_{2}),\nonumber \\
x_{-}(1,2) & =x_{1}\theta(-x_{1}+x_{2})+x_{2}\theta(x_{1}-x_{2}),\nonumber \\
y_{+}(1,2) & =y_{1}\theta(y_{1}-y_{2})+y_{2}\theta(-y_{1}+y_{2}),\nonumber \\
y_{-}(1,2) & =y_{1}\theta(-y_{1}+y_{2})+y_{2}\theta(y_{1}-y_{2}).
\end{align}
Inserting the above formula into $I_{2}^{(0)}$ and using the equation
\begin{equation}
x\theta(x)=\frac{x+|x|}{2},
\end{equation}
we obtain
\begin{eqnarray}
I_{2}^{(0)} & = & -(x_{1}-x_{2})(y_{1}-y_{2})\left[\theta(x_{1}-x_{2})\theta(y_{1}-y_{2})+\theta(-x_{1}+x_{2})\theta(-y_{1}+y_{2})\right]\nonumber \\
 & = & -\frac{1}{2}\left[(x_{1}-x_{2})(y_{1}-y_{2})+\left|(x_{1}-x_{2})(y_{1}-y_{2})\right|\right].\label{eq:ch0}
\end{eqnarray}
This form will be used in the next section.

\subsubsection{$I_{3322}$}

Like $I_{2}$ in (\ref{eq:I2}), we can define $I_{3}$ as
\begin{eqnarray}
I_{3} & = & x_{1}y_{2}+x_{1}y_{3}+x_{2}y_{1}-x_{2}y_{2}+x_{2}y_{3}+x_{3}y_{1}+x_{3}y_{2}-x_{3}y_{3}\nonumber \\
 &  & -(x_{1}+x_{2})B-A(y_{1}+y_{2}),\label{eq:i3-symm}
\end{eqnarray}
which involves elements of sets $a$ and $b$ in (\ref{eq:s-eq})
with $m=n=3$. We now prove the $I_{3322}$ inequality, $I_{3}\leqslant0$.
We use boundary conditions (\ref{eq:bc}), or, explicitly,
\begin{align}
0\leqslant x_{-}(1,2,3)\leqslant x_{1},x_{2},x_{3} & \leqslant x_{+}(1,2,3)\leqslant A,\nonumber \\
0\leqslant y_{-}(1,2,3)\leqslant y_{1},y_{2},y_{3} & \leqslant y_{+}(1,2,3)\leqslant B,
\end{align}
and replace $A$ and $B$ in (\ref{eq:i3-symm}) with $x_{+}(1,2,3)$
and $y_{+}(1,2,3)$, respectively, to obtain
\begin{eqnarray}
I_{3} & \leqslant & x_{1}(y_{2}+y_{3})+x_{2}(y_{1}+y_{3})+x_{3}(y_{1}+y_{2})-x_{2}y_{2}-x_{3}y_{3}\nonumber \\
 &  & -(x_{1}+x_{2})y_{+}(1,2,3)-x_{+}(1,2,3)(y_{1}+y_{2}).\label{eq:I3-e-1}
\end{eqnarray}
Using the inequalities
\begin{align}
-x_{2}y_{2} & \leqslant-x_{2}y_{-}(1,2,3)-x_{-}(1,2,3)y_{2}+x_{-}(1,2,3)y_{-}(1,2,3),\nonumber \\
-x_{3}y_{3} & \leqslant-x_{3}y_{+}(1,2,3)-x_{+}(1,2,3)y_{3}+x_{+}(1,2,3)y_{+}(1,2,3),\label{eq:I3-ee}
\end{align}
we can enlarge the right-hand side of (\ref{eq:I3-e-1}) to obtain
\begin{eqnarray}
I_{3} & \leqslant & x_{1}(y_{2}+y_{3})+x_{2}(y_{1}+y_{3})+x_{3}(y_{1}+y_{2})\nonumber \\
 &  & -x_{2}y_{-}(1,2,3)-x_{-}(1,2,3)y_{2}+x_{-}(1,2,3)y_{-}(1,2,3)\nonumber \\
 &  & -x_{3}y_{+}(1,2,3)-x_{+}(1,2,3)y_{3}+x_{+}(1,2,3)y_{+}(1,2,3)\nonumber \\
 &  & -(x_{1}+x_{2})y_{+}(1,2,3)-x_{+}(1,2,3)(y_{1}+y_{2}).\label{eq:I3-e-2}
\end{eqnarray}
The inequalities in (\ref{eq:I3-ee}) are true because they can be
put into the following form:
\begin{align}
-\left[x_{2}-x_{-}(1,2,3)\right]\left[y_{2}-y_{-}(1,2,3)\right] & \leqslant0,\nonumber \\
-\left[x_{3}-x_{+}(1,2,3)\right]\left[y_{3}-y_{+}(1,2,3)\right] & \leqslant0.
\end{align}
After adding additional positive terms to enlarge the right-hand side
of (\ref{eq:I3-e-2}), we can derive
\begin{eqnarray}
I_{3} & \leqslant & x_{1}(y_{2}+y_{3})+x_{2}(y_{1}+y_{3})+x_{3}(y_{1}+y_{2})\nonumber \\
 &  & -(x_{1}+x_{2}+x_{3})y_{+}(1,2,3)-x_{+}(1,2,3)(y_{1}+y_{2}+y_{3})\nonumber \\
 &  & -\left[x_{1}-x_{+}(1,2,3)\right]y_{-}(1,2,3)-x_{2}y_{-}(1,2,3)-\left[x_{3}-x_{+}(1,2,3)\right]y_{-}(1,2,3)\nonumber \\
 &  & -x_{-}(1,2,3)\left[y_{1}-y_{+}(1,2,3)\right]-x_{-}(1,2,3)y_{2}-x_{-}(1,2,3)\left[y_{3}-y_{+}(1,2,3)\right]\nonumber \\
 &  & +x_{-}(1,2,3)y_{-}(1,2,3)+x_{+}(1,2,3)y_{+}(1,2,3)\nonumber \\
 & = & -(x_{1}y_{1}+x_{2}y_{2}+x_{3}y_{3})+x_{+}(1,2,3)y_{-}(1,2,3)+x_{-}(1,2,3)y_{+}(1,2,3)\nonumber \\
 &  & +\left[x_{1}+x_{2}+x_{3}-x_{+}(1,2,3)-x_{-}(1,2,3)\right]\nonumber \\
 &  & \times\left[y_{1}+y_{2}+y_{3}-y_{+}(1,2,3)-y_{-}(1,2,3)\right].\label{eq:I3-e-3}
\end{eqnarray}
Finally, we define $I_{3}^{(0)}$ as the last polynomial of (\ref{eq:I3-e-3})
for later use,
\begin{eqnarray}
I_{3}^{(0)} & = & -(x_{1}y_{1}+x_{2}y_{2}+x_{3}y_{3})+x_{+}(1,2,3)y_{-}(1,2,3)\nonumber \\
 &  & +x_{-}(1,2,3)y_{+}(1,2,3)+x_{r}y_{r},\label{eq:I3-0}
\end{eqnarray}
where $x_{r}$ and $y_{r}$ denote the middle terms in sets $a$ and
$b$ as
\begin{align}
x_{r} & =x_{1}+x_{2}+x_{3}-x_{+}(1,2,3)-x_{-}(1,2,3),\nonumber \\
y_{r} & =y_{1}+y_{2}+y_{3}-y_{+}(1,2,3)-y_{-}(1,2,3).
\end{align}
It is obvious that $I_{3}^{(0)}$ in Eq. (\ref{eq:I3-0}) is the difference
between the reversed sum and the unordered sum for sets $a$ and $b$
in (\ref{eq:s-eq}) with $m=n=3$. From the rearrangement inequality,
we have
\begin{equation}
I_{3}\leqslant I_{3}^{(0)}\leqslant0.
\end{equation}
This concludes the proof of the $I_{3322}$ inequality using the method
of the rearrangement inequality.

\subsection{Method of linear inequality}

\label{subsec:method-of-linear}In Sec. \ref{subsec:method-of-rearrangement},
we derived two algebraic inequalities in connection with low-order
Bell-CH and rearrangement inequalities. But it is not easy to generalize
this method to higher-order cases. In Ref. \citep{Collins2004}, the
authors obtained a specific type of Bell-CH inequalities and found
the relationship between lower-order and higher-order inequalities
of this type. Following Ref. \citep{Collins2004}, there has been
a lot of discussion on the perspective of mathematics along this line
\citep{Avis2005,Avis2006,Avis2007,Avis2008}. Inspired by these works,
we find a general method which we call the method of linear inequality
to prove higher-order inequalities $I_{mn22}$ from the lowest-order
one systematically. As a by-product, this method can be used to construct
higher-order inequalities more effectively than the traditional method.
We will prove a lemma using the property of monotonicity. Then we
will apply the lemma to linear functions and obtain a theorem. Finally,
we will use this method to prove a large variety of inequalities that
have been obtained.

We consider two sets $a$ and $b$ defined by Eq. (\ref{eq:s-eq})
with conditions (\ref{eq:bc}). Starting from positivity inequalities
\begin{eqnarray}
-x_{i}y_{j} & \leqslant & 0,\nonumber \\
x_{i}(y_{j}-B) & \leqslant & 0,\nonumber \\
(x_{i}-A)y_{j} & \leqslant & 0,\label{eq:positivity}
\end{eqnarray}
we present a theorem for a specific type of algebraic Bell-CH inequalities.

\emph{Theorem 1. }$I_{kk}(x_{1},\cdots,x_{k}|y_{1},\cdots,y_{k})\leqslant0$
is an algebraic Bell-CH inequality for $k\geqslant2$ with variables
$x_{i}$ and $y_{j}$ in sets $a$ and $b$, respectively, where $I_{kk}(x_{1},\cdots,x_{k}|y_{1},\cdots,y_{k})$
is defined as
\begin{equation}
I_{kk}(x_{1},\cdots,x_{k}|y_{1},\cdots,y_{k})=\sum_{j=1}^{k}\sum_{i=1}^{k+1-j}x_{i}y_{j}-\sum_{i=2}^{k}x_{i}y_{k+2-i}-\sum_{i=1}^{k-1}(k-i)x_{i}B-Ay_{1}.\label{eq:ikk-theorem}
\end{equation}

\emph{Proof. }Let $f(x_{i},\{c_{j}^{y}\})=x_{i}(\sum_{j}c_{j}^{y}y_{j})$
and $g(y_{i},\{c_{j}^{x}\})=(\sum_{j}c_{j}^{x}x_{j})y_{i}$, where
$x_{i}$ and $y_{j}$ are variables in sets $a$ and $b$ and $c_{j}^{x}$
and $c_{j}^{y}$ are coefficients. Then we have the following linear
inequalities:
\begin{eqnarray}
f(x_{i},\{c_{j}^{y}\}) & \leqslant & \theta(\sum_{j}c_{j}^{y}y_{j})f(A,\{c_{j}^{y}\})+\theta(-\sum_{j}c_{j}^{y}y_{j})f(0,\{c_{j}^{y}\}),\nonumber \\
g(y_{i},\{c_{j}^{x}\}) & \leqslant & \theta(\sum_{j}c_{j}^{x}x_{j})g(B,\{c_{j}^{x}\})+\theta(-\sum_{j}c_{j}^{x}x_{j})g(0,\{c_{j}^{x}\}).\label{eq:co-ieq}
\end{eqnarray}
We can prove $I_{kk}(x_{1},\cdots,x_{k}|y_{1},\cdots,y_{k})\leqslant0$
with the help of (\ref{eq:co-ieq}) by the induction method. First,
we should prove the case for $k=2$, which is just the algebraic version
of the original Bell-CH inequality\emph{. }We have
\begin{eqnarray}
I_{22}(x_{1},x_{2}|y_{1},y_{2}) & = & (x_{1}+x_{2})y_{1}+(x_{1}-x_{2})y_{2}-x_{1}B-Ay_{1}\nonumber \\
 & \leqslant & \theta(x_{1}-x_{2})I_{22}(x_{1},x_{2}|y_{1},B)+\theta(-x_{1}+x_{2})I_{22}(x_{1},x_{2}|y_{1},0)\nonumber \\
 & = & \theta(x_{1}-x_{2})[(x_{1}+x_{2})y_{1}-x_{2}B-Ay_{1}]\nonumber \\
 &  & +\theta(-x_{1}+x_{2})[(x_{1}+x_{2})y_{1}-x_{1}B-Ay_{1}],
\end{eqnarray}
where we have used (\ref{eq:co-ieq}) for $g(y_{2},\{c_{j}^{x}\})=(x_{1}-x_{2})y_{2}$
with $\{c_{j}^{x}\}=\{1,-1\}$. Then we can rewrite the right-hand
side of the above inequality and obtain
\begin{eqnarray}
I_{22}(x_{1},x_{2}|y_{1},y_{2}) & \leqslant & \theta(x_{1}-x_{2})[(x_{1}-A)y_{1}+x_{2}(y_{1}-B)]\nonumber \\
 &  & +\theta(-x_{1}+x_{2})[(x_{2}-A)y_{1}+x_{1}(y_{1}-B)]\nonumber \\
 & \leqslant & 0.\label{eq:nch-c}
\end{eqnarray}
Thus, the inequality holds for $k=2$.

Then we assume the inequality holds for $k-1$ with
\begin{eqnarray}
I_{k-1,k-1}(x_{1},x_{3},\cdots,x_{k}|y_{1},\cdots,y_{k-1}) & \leqslant & 0,\nonumber \\
I_{k-1,k-1}(x_{2},x_{3},\cdots,x_{k}|y_{1},\cdots,y_{k-1}) & \leqslant & 0.
\end{eqnarray}
According to the induction method, the inequality should hold for
$I_{kk}(x_{1},\cdots,x_{k}|y_{1},\cdots,y_{k})$ in (\ref{eq:ikk-theorem}).
We now apply (\ref{eq:co-ieq}) to $I_{kk}$ for $g(y_{k},\{c_{j}^{x}\})=(x_{1}-x_{2})y_{k}$,
with $\{c_{j}^{x},j=1,2,\cdots,k\}=\{1,-1,0,\cdots,0\}$, as
\begin{eqnarray}
 &  & I_{kk}(x_{1},\cdots,x_{k}|y_{1},\cdots,y_{k})\nonumber \\
 & \leqslant & \theta(x_{1}-x_{2})I_{kk}(x_{1},\cdots,x_{k}|y_{1},\cdots y_{k-1},B)+\theta(-x_{1}+x_{2})I_{kk}(x_{1},\cdots,x_{k}|y_{1},\cdots,y_{k-1},0)\nonumber \\
 & = & \theta(x_{1}-x_{2})[(\sum_{i=1}^{k}x_{i})y_{1}+(\sum_{i=1}^{k-1}x_{i}-x_{k})y_{2}+\cdots+(x_{1}+x_{2}-x_{3})y_{k-1}\nonumber \\
 &  & +(x_{1}-x_{2})B-\sum_{i=1}^{k-1}(k-i)x_{i}B-Ay_{1}]\nonumber \\
 &  & +\theta(-x_{1}+x_{2})[(\sum_{i=1}^{k}x_{i})y_{1}+(\sum_{i=1}^{k-1}x_{i}-x_{k})y_{2}+\cdots+(x_{1}+x_{2}-x_{3})y_{k-1}\nonumber \\
 &  & -\sum_{i=1}^{k-1}(k-i)x_{i}B-Ay_{1}].
\end{eqnarray}
We can then rewrite the right-hand side of the above inequality and
obtain
\begin{eqnarray}
 &  & I_{kk}(x_{1},\cdots,x_{k}|y_{1},\cdots,y_{k})\nonumber \\
 & \leqslant & \theta(x_{1}-x_{2})\{I_{k-1,k-1}(x_{1},x_{3},\cdots,x_{k}|y_{1},\cdots,y_{k-1})+\sum_{i=1}^{k-1}[x_{2}(y_{i}-B)]\}\nonumber \\
 &  & +\theta(-x_{1}+x_{2})\{I_{k-1,k-1}(x_{2},x_{3},\cdots,x_{k}|y_{1},\cdots,y_{k-1})+\sum_{i=1}^{k-1}[x_{1}(y_{i}-B)]\}\nonumber \\
 & \leqslant & I_{kk}^{(0)}(x_{1},\cdots,x_{k}|y_{1},\cdots,y_{k})\nonumber \\
 & \leqslant & 0.\label{eq:ch-mn}
\end{eqnarray}
Here, $I_{kk}^{(0)}$ is given by
\begin{eqnarray}
 &  & I_{kk}^{(0)}(x_{1},\cdots,x_{k}|y_{1},\cdots,y_{k})\nonumber \\
 & = & \max\{I_{k-1,k-1}(x_{1},x_{3},\cdots,x_{k}|y_{1},\cdots,y_{k-1})+\sum_{i=1}^{k-1}[x_{2}(y_{i}-B)],\nonumber \\
 &  & I_{k-1,k-1}(x_{2},x_{3},\cdots,x_{k}|y_{1},\cdots,y_{k-1})+\sum_{i=1}^{k-1}[x_{1}(y_{i}-B)]\}.
\end{eqnarray}
So we prove in (\ref{eq:ch-mn}) that the inequality holds for $k$.
This concludes the proof of the theorem.

The same method can also be applied to prove general inequalities
$I_{mn22}$ recursively. For example, we can prove $I_{3}$ in (\ref{eq:i3-symm})
as the symmetric version of $I_{33}(x_{1},x_{2},x_{3}|y_{1},y_{2},y_{3})$:
$I_{3}$ can be obtained from $I_{33}(x_{1},x_{2},x_{3}|y_{1},y_{2},y_{3})$
by the transformation $x_{1}\rightarrow A-x_{1}$, $y_{2}\rightarrow B-y_{2}$,
$y_{3}\rightarrow B-y_{3}$ and then relabeling indices of $\{x_{i}\}$
and $\{y_{j}\}$ by interchanging $1\leftrightarrow3$. Hence, $I_{3}\leqslant0$
can be proved by using lower-order inequalities with the help of (\ref{eq:co-ieq}).

As another example, we can reconstruct and prove $I_{5322}$ by applying
linear inequalities. We use the polynomial $I_{53}(x_{1},x_{2},x_{3},x_{4},x_{5}|y_{1},y_{2},y_{3})$
to represent $I_{5322}$ as
\begin{eqnarray}
 &  & I_{53}(x_{1},x_{2},x_{3},x_{4},x_{5}|y_{1},y_{2},y_{3})\nonumber \\
 & \equiv & x_{1}y_{1}-x_{1}y_{2}+x_{1}y_{3}+x_{2}y_{2}+x_{2}y_{3}+x_{3}y_{1}+x_{3}y_{2}+x_{4}y_{1}-x_{4}y_{3}\nonumber \\
 &  & -x_{5}y_{1}+x_{5}y_{2}-x_{5}y_{3}-(x_{1}+x_{2}+x_{3})B-A(y_{1}+y_{2})\nonumber \\
 & \leqslant & 0.\label{eq:ieq-53}
\end{eqnarray}
To prove inequality (\ref{eq:ieq-53}), we can apply linear inequalities
(\ref{eq:co-ieq}) for $f(x_{4},\{c_{j}^{y}\})=x_{4}(y_{1}-y_{3})$,
with $\{c_{j}^{y},j=1,2,3\}=\{1,0,-1\}$, and obtain
\begin{eqnarray}
 &  & I_{53}(x_{1},x_{2},x_{3},x_{4},x_{5}|y_{1},y_{2},y_{3})\nonumber \\
 & \leqslant & \theta(y_{1}-y_{3})I_{53}(x_{1},x_{2},x_{3},A,x_{5}|y_{1},y_{2},y_{3})+\theta(-y_{1}+y_{3})I_{53}(x_{1},x_{2},x_{3},0,x_{5}|y_{1},y_{2},y_{3})\nonumber \\
 & = & \theta(y_{1}-y_{3})[I_{22}(x_{2},x_{1}|y_{3},y_{2})+I_{22}(x_{3},x_{5}|y_{2},y_{1})+x_{1}(y_{1}-B)-x_{5}y_{3}]\nonumber \\
 &  & +\theta(-y_{1}+y_{3})[I_{22}(x_{2},x_{5}|y_{2},y_{3})+I_{22}(x_{3},x_{1}|y_{1},y_{2})+x_{1}(y_{3}-B)-x_{5}y_{1}]\nonumber \\
 & \leqslant & 0,
\end{eqnarray}
where we have used inequality (\ref{eq:nch-c}) for $I_{22}(x_{2},x_{1}|y_{3},y_{2})$,
$I_{22}(x_{3},x_{5}|y_{2},y_{1})$, $I_{22}(x_{2},x_{5}|y_{2},y_{3})$,
and $I_{22}(x_{3},x_{1}|y_{1},y_{2})$. It was shown in Refs. \citep{Deza2016,Deza2020}
that there is only one $I5322$, and the explicit form was first found
by \citep{Quintino2014}. Here, we give the proof of the corresponding
algebraic one using the linear inequality method.

In the Supplemental Material \citep{Supplemental}, we summarize proofs
of 257 algebraic Bell-CH inequalities using the linear inequality
method. In these proofs, we can show that all inequalities can be
reduced to second- and third-order ones.

\section{Bell-CH inequalities for binary settings}

\label{sec:Bell-CH}According to ``objective local theories'' introduced
by Clauser and Horne \citep{Clauser1974}, any physical system, either
a classical or quantum-mechanical one, can be considered as a state.
The state is labeled by some local hidden variables $\lambda$ without
any other assumptions. For example, in a bipartite correlated system,
$P(x,\lambda)$ describes the probability density of some certain
measurement outcome $x$ in one subsystem, and $P(x,y,\lambda)$ is
the correlation probability density of measurement outcomes $x$ and
$y$ in two subsystems.

From the definition of the state and local hidden variables, we can
derive Bell-CH inequalities from algebraic ones. In a theory of local
hidden variables, the physically detectable probability density $P(x)$
is related to a hidden variable $\lambda$, which is assumed to be
drawn from the probability distribution $\rho(\lambda)\in[0,1]$ as
\begin{equation}
P(x)=\int P(x,\lambda)\rho(\lambda)d\lambda.
\end{equation}
Likewise, the joint probability density $P(x,y)$ is obtained with
\begin{equation}
P(x,y)=\int P(x,\lambda)P(y,\lambda)\rho(\lambda)d\lambda.
\end{equation}
Let us take the derivation of the CH inequality as an example. From
the $I_{2222}$ inequality, replacing $x_{i}/A$ and $y_{i}/B$ with
$P(x_{i},\lambda)$ and $P(y_{i},\lambda)$ for $i=1,2$, respectively,
we obtain
\begin{eqnarray}
P(x_{1},\lambda)\left[P(y_{1},\lambda)+P(y_{2},\lambda)\right]+P(x_{2},\lambda)\left[P(y_{1},\lambda)-P(y_{2},\lambda)\right]-P(x_{1},\lambda)-P(y_{1},\lambda) & \leqslant & 0.\nonumber \\
\end{eqnarray}
Multiplying the above inequality by $\rho(\lambda)$ and taking an
integration over $\lambda$, we obtain the CH inequality
\begin{equation}
I_{2,CH}\leqslant0,\label{eq:ch}
\end{equation}
where $I_{2,CH}$ is defined as
\begin{eqnarray}
I_{2,CH} & = & P(x_{1},y_{1})+P(x_{1},y_{2})+P(x_{2},y_{1})-P(x_{2},y_{2})-P(x_{1})-P(y_{1}).
\end{eqnarray}

In correspondence to (\ref{eq:I2-e-r}), we obtain the inequality
\begin{equation}
I_{2,CH}\leqslant I_{2,CH}^{(0)}\leqslant0,\label{eq:nieq-1}
\end{equation}
where $I_{2,CH}^{(0)}$ is defined as
\begin{eqnarray}
I_{2,CH}^{(0)} & = & -\frac{1}{2}\int\left\{ \left[P(x_{1},\lambda)-P(x_{2},\lambda)\right]\left[P(y_{1},\lambda)-P(y_{2},\lambda)\right]\right.\nonumber \\
 &  & \left.+\left|\left[P(x_{1},\lambda)-P(x_{2},\lambda)\right]\left[P(y_{1},\lambda)-P(y_{2},\lambda)\right]\right|\right\} \rho(\lambda)d\lambda.
\end{eqnarray}
Here, we use Eq. (\ref{eq:ch0}) and make the replacements $x_{i}/A\rightarrow P(x_{i},\lambda)$
and $y_{i}/B\rightarrow P(y_{i},\lambda)$ for $i=1,2$, respectively,
multiplied by $\rho(\lambda)$ and integrated over $\lambda$. We
apply Jensen's inequality in Appendix \ref{sec:jensen} assuming the
convex function $\varphi(x)=\left|x\right|$, which leads to an upper
bound for $I_{2,CH}^{(0)}$ and then for $I_{2,CH}$,
\begin{eqnarray}
I_{2,CH}\leqslant I_{2,CH}^{(0)} & \leqslant & -\frac{1}{2}\left\{ \left[P(x_{1},y_{1})-P(x_{1},y_{2})-P(x_{2},y_{1})+P(x_{2},y_{2})\right]\right.\nonumber \\
 &  & \left.+\left|P(x_{1},y_{1})-P(x_{1},y_{2})-P(x_{2},y_{1})+P(x_{2},y_{2})\right|\right\} ,\label{eq:nch}
\end{eqnarray}
where we have employed inequality (\ref{eq:nieq-1}). We can easily
see that the upper bound of $I_{2,CH}^{(0)}$ is less than or equal
to zero. 

Another approach to our class of inequalities is through the method
of linear inequalities in deriving algebraic inequalities $I_{mn22}$.
We take a type of Bell-CH inequality as an example. In correspondence
to inequality (\ref{eq:ch-mn}), we obtain the inequality
\begin{equation}
I_{kk;Q}\leqslant I_{kk;Q}^{(0)}\leqslant0,\label{eq:nieq-2}
\end{equation}
in which we have defined
\begin{eqnarray}
I_{kk;Q} & \equiv & (AB)^{-1}\int d\lambda\rho(\lambda)I_{kk}\left[AP(x_{1},\lambda),\cdots,AP(x_{k},\lambda)|BP(y_{1},\lambda),\cdots,BP(y_{k},\lambda)\right],\nonumber \\
I_{kk;Q}^{(0)} & \equiv & (AB)^{-1}\int d\lambda\rho(\lambda)I_{kk}^{(0)}\left[AP(x_{1},\lambda),\cdots,AP(x_{k},\lambda)|BP(y_{1},\lambda),\cdots,BP(y_{k},\lambda)\right].\nonumber \\
\end{eqnarray}
Here, we have made the replacements in (\ref{eq:ch-mn}) $x_{i}/A\rightarrow P(x_{i},\lambda)$
and $y_{i}/B\rightarrow P(y_{i},\lambda)$ for $i=1,\ldots,k$, respectively,
multiplied by $\rho(\lambda)$ and integrated over $\lambda$. The
inequality $I_{kk,Q}^{(0)}\leqslant0$ leads to the following alternative
inequality:
\begin{eqnarray}
\mathrm{max}\left\{ I_{k-1,k-1;Q}^{(1)}+\sum_{i=1}^{k-1}\left[P(x_{2},y_{i})-P(x_{2})\right],I_{k-1,k-1;Q}^{(2)}+\sum_{i=1}^{k-1}\left[P(x_{1},y_{i})-P(x_{1})\right]\right\}  & \leqslant & 0,\nonumber \\
\label{eq:nch-m}
\end{eqnarray}
where $I_{k-1,k-1;Q}^{(1)}$ and $I_{k-1,k-1;Q}^{(2)}$ are defined
as
\begin{eqnarray}
I_{k-1,k-1;Q}^{(1)} & \equiv & \sum_{j=1}^{k-1}\sum_{i=1,i\neq2}^{k-j}P(x_{i},y_{j})-\sum_{i=3}^{k-1}P(x_{i},y_{k+1-i})-\sum_{i=1,i\neq2}^{k-2}(k-1-i)P(x_{i})-P(y_{1}),\nonumber \\
I_{k-1,k-1;Q}^{(2)} & \equiv & \sum_{j=1}^{k-1}\sum_{i=2}^{k-j}P(x_{i},y_{j})-\sum_{i=3}^{k-1}P(x_{i},y_{k+1-i})-\sum_{i=2}^{k-2}(k-1-i)P(x_{i})-P(y_{1}).
\end{eqnarray}
Note that this inequality can be tested in physical systems.

Applying this method to $Imn22$, we can obtain a set of Bell-CH inequalities
which is summarized in the Supplemental Material \citep{Supplemental}.
In the next section, we will test our inequalities with special quantum
entangled states.

\section{Testing our Bell-CH inequalities with quantum entangled states}

\label{sec:testing}To test our class of Bell-CH inequalities we consider
a simple quantum system with two qubits. The entangled states of two
qubits we will use in the test are parameterized as
\begin{equation}
\left|\psi(\theta)\right\rangle =\cos\theta\left|00\right\rangle +\sin\theta\left|11\right\rangle .\label{eq:t-state}
\end{equation}
For such quantum states, the correlation probabilities can be obtained
from expectation values of operators acting on the Hilbert space,
\begin{eqnarray}
P(x_{i}) & = & \left\langle \psi(\theta)\right|x_{i}\varotimes I_{y}\left|\psi(\theta)\right\rangle ,\nonumber \\
P(y_{j}) & = & \left\langle \psi(\theta)\right|I_{x}\varotimes y_{j}\left|\psi(\theta)\right\rangle ,\nonumber \\
P(x_{i},y_{j}) & = & \left\langle \psi(\theta)\right|x_{i}\varotimes y_{j}\left|\psi(\theta)\right\rangle ,
\end{eqnarray}
with $x_{i}$ $(y_{j})$ being the projectors measured by Alice (Bob)
on $x$ ($y$), and $I_{x}$ $(I_{y})$ being the unit operators acting
on $x$ ($y$).

For entangled states (\ref{eq:t-state}), we define $Q$, $Q_{a}$,
and $Q_{b}$ as the maximum violations of $Imn22$ and the corresponding
Bell-CH inequalities $Imn22^{a}$ and $Imn22^{b}$ presented in the
Supplemental Material \citep{Supplemental}, respectively. For the
test, we have calculated the maximum violations of our inequalities
for all $Imn22$ listed in \citep{Supplemental}. For states with
the maximum violation characterized by $\theta_{max}$, we have also
calculated the resistance to noise $\lambda_{max}$ defined through
the mixture state
\begin{equation}
\rho=\lambda\left|\psi(\theta_{max})\right\rangle \left\langle \psi(\theta_{max})\right|+(1-\lambda)\frac{I}{4},
\end{equation}
so that it does not violate the inequality marginally, where $1-\lambda$
is the parameter for white noise. In Table 1, we give some selected
results of 32 $Imn22$ and our corresponding Bell-CH inequalities
from the complete table in the Supplemental Material \citep{Supplemental}.

\begin{longtable}[c]{>{\centering}m{1.5cm}>{\centering}m{1.5cm}>{\centering}m{1.5cm}>{\centering}m{1.5cm}>{\centering}m{1.5cm}>{\centering}m{1.5cm}>{\centering}m{1.5cm}>{\centering}m{1.5cm}>{\centering}m{1.5cm}>{\centering}m{1.5cm}}
\caption{Maximum violations $Q$, $Q_{a}$, and $Q_{b}$; related parameters
$\theta_{max}/\pi$, $\theta_{max}^{a}/\pi$, and $\theta_{max}^{b}/\pi$;
and resistances to noise $\lambda_{max}$, $\lambda_{max}^{a}$, and
$\lambda_{max}^{b}$ of 32 $Imn22$ (here we use the original names
in the literature to represent them) and corresponding Bell-CH inequalities
$Imn22^{a}$ and $Imn22^{b}$.}
\tabularnewline
\midrule 
Name &
$Q$ &
$\theta_{max}/\pi$ &
$\lambda_{max}$ &
$Q_{a}$ &
$\theta_{max}^{a}/\pi$ &
$\lambda_{max}^{a}$ &
$Q_{b}$ &
$\theta_{max}^{b}/\pi$ &
$\lambda_{max}^{b}$\tabularnewline
\midrule
\endfirsthead
\caption{(Continued.)}
\tabularnewline
\midrule 
Name &
$Q$ &
$\theta_{max}/\pi$ &
$\lambda_{max}$ &
$Q_{a}$ &
$\theta_{max}^{a}/\pi$ &
$\lambda_{max}^{a}$ &
$Q_{b}$ &
$\theta_{max}^{b}/\pi$ &
$\lambda_{max}^{b}$\tabularnewline
\midrule
\endhead
\midrule 
$I_{3322}$ &
0.25 &
0.25 &
0.8 &
0.2071 &
0.25 &
0.7836 &
- &
- &
-\tabularnewline
\midrule 
$I_{4422}^{4}$ &
0.056 &
0.1316 &
0.9728 &
0.2361 &
0.2332 &
0.864 &
0.4142 &
0.25 &
0.7071\tabularnewline
\midrule 
$I_{4422}^{14}$ &
0.4103 &
0.238 &
0.8298 &
0.4282 &
0.2377 &
0.8034 &
0.3793 &
0.2304 &
0.7981\tabularnewline
\midrule 
$I_{4422}^{16}$ &
0.2407 &
0.219 &
0.8791 &
0.226 &
0.2362 &
0.8691 &
0.2071 &
0.25 &
0.8579\tabularnewline
\midrule 
$I_{4422}^{18}$ &
0.1812 &
0.168 &
0.9508 &
0.2983 &
0.2195 &
0.9096 &
0.5436 &
0.2278 &
0.7863\tabularnewline
\midrule 
$J_{4422}^{3}$ &
0.7249 &
0.2448 &
0.838 &
0.7337 &
0.2419 &
0.8267 &
- &
- &
-\tabularnewline
\midrule 
$J_{4422}^{4}$ &
0.6862 &
0.2452 &
0.8138 &
0.6579 &
0.2426 &
0.7917 &
- &
- &
-\tabularnewline
\midrule 
$J_{4422}^{5}$ &
0.5007 &
0.2404 &
0.8749 &
0.4434 &
0.2303 &
0.8494 &
- &
- &
-\tabularnewline
\midrule 
$J_{4422}^{7}$ &
0.3642 &
0.2304 &
0.8917 &
0.6057 &
0.2376 &
0.7675 &
- &
- &
-\tabularnewline
\midrule 
$J_{4422}^{8}$ &
0.2657 &
0.19 &
0.9186 &
0.2772 &
0.2057 &
0.9002 &
- &
- &
-\tabularnewline
\midrule 
$J_{4422}^{12}$ &
0.4198 &
0.25 &
0.9147 &
0.5797 &
0.2407 &
0.8381 &
- &
- &
-\tabularnewline
\midrule 
$J_{4422}^{13}$ &
0.5629 &
0.2422 &
0.8766 &
0.5541 &
0.2213 &
0.8633 &
0.636 &
0.2421 &
0.8251\tabularnewline
\midrule 
$J_{4422}^{15}$ &
0.6133 &
0.2433 &
0.8412 &
0.5668 &
0.2416 &
0.8411 &
- &
- &
-\tabularnewline
\midrule 
$J_{4422}^{21}$ &
0.5441 &
0.2397 &
0.8655 &
0.5717 &
0.2255 &
0.8279 &
- &
- &
-\tabularnewline
\midrule 
$J_{4422}^{30}$ &
0.2459 &
0.1832 &
0.9242 &
0.4201 &
0.2398 &
0.8264 &
- &
- &
-\tabularnewline
\midrule 
$J_{4422}^{31}$ &
0.2133 &
0.1847 &
0.9336 &
0.3796 &
0.2432 &
0.8556 &
- &
- &
-\tabularnewline
\midrule 
$J_{4422}^{34}$ &
0.4075 &
0.2377 &
0.8804 &
0.3925 &
0.2318 &
0.8515 &
- &
- &
-\tabularnewline
\midrule 
$J_{4422}^{42}$ &
0.6012 &
0.2362 &
0.8331 &
0.6722 &
0.2462 &
0.7881 &
- &
- &
-\tabularnewline
\midrule 
$J_{4422}^{51}$ &
0.6678 &
0.2419 &
0.8046 &
0.675 &
0.2466 &
0.7874 &
- &
- &
-\tabularnewline
\midrule 
$J_{4422}^{88}$ &
0.616 &
0.2468 &
0.7851 &
0.5956 &
0.2446 &
0.7705 &
- &
- &
-\tabularnewline
\midrule 
$J_{4422}^{113}$ &
0.8196 &
0.2399 &
0.83 &
0.8484 &
0.2448 &
0.793 &
- &
- &
-\tabularnewline
\midrule 
$N_{4422}^{6}$ &
0.5972 &
0.2496 &
0.8543 &
0.5695 &
0.2426 &
0.8404 &
- &
- &
-\tabularnewline
\midrule 
$N_{4422}^{9}$ &
0.7399 &
0.2445 &
0.8352 &
0.8259 &
0.2404 &
0.7841 &
- &
- &
-\tabularnewline
\midrule 
$N_{4422}^{10}$ &
0.5 &
0.25 &
0.8333 &
0.4331 &
0.2449 &
0.822 &
- &
- &
-\tabularnewline
\midrule 
$A_{10}$ &
0.4154 &
0.229 &
0.8082 &
0.3944 &
0.2388 &
0.7918 &
- &
- &
-\tabularnewline
\midrule 
$A_{11}$ &
0.4561 &
0.2379 &
0.7933 &
0.3944 &
0.2388 &
0.7918 &
- &
- &
-\tabularnewline
\midrule 
$A_{13}$ &
0.4031 &
0.2375 &
0.8128 &
0.4142 &
0.25 &
0.7836 &
0.4142 &
0.25 &
0.7836\tabularnewline
\midrule 
$A_{16}$ &
0.416 &
0.2402 &
0.8278 &
0.4353 &
0.2447 &
0.7751 &
0.3944 &
0.2388 &
0.7601\tabularnewline
\midrule 
$A_{34}$ &
0.514 &
0.2461 &
0.7956 &
0.535 &
0.2476 &
0.7659 &
- &
- &
-\tabularnewline
\midrule 
$A_{69}$ &
0.3304 &
0.2245 &
0.8833 &
0.4459 &
0.2393 &
0.8177 &
0.5148 &
0.25 &
0.7727\tabularnewline
\midrule 
$A_{83}$ &
0.6962 &
0.2438 &
0.798 &
0.62 &
0.2424 &
0.784 &
- &
- &
-\tabularnewline
\midrule 
$A_{88}$ &
0.0768 &
0.1575 &
0.9702 &
0.197 &
0.2356 &
0.9103 &
0.25 &
0.25 &
0.8571\tabularnewline
\bottomrule
\end{longtable}

For all $Imn22$, we find that our Bell-CH inequalities have positive
violation by entangled states (\ref{eq:t-state}). According to the
preceding section, our inequalities can be decomposed into groups
of inequalities formed by combinations of some original low-order
$Imn22$; the nonlocality implied by the violation of high-order $Imn22$
can be replaced by that of our inequalities. Moreover, the resistance
to noise for some of these inequalities is lower than that of the
original ones; hence, our inequalities could be better candidates
for testing nonlocality in some physical systems. 

\section{Summary and conclusions}

\label{sec:summary}In this work effective methods for the rearrangement
and linear inequality were employed to prove the Bell-CH inequalities.
Alternative types of Bell-CH inequalities were found to be violated
by entangled states. The main results are summarized as follows. First,
a large variety of $Imn22$ ($m,n\le5$) can be easily derived through
the rearrangement inequality and the linear inequality. Second, along
with $I2222$ or the CH inequality, we gave an inequality in (\ref{eq:nch})
using the rearrangement inequality method. Third, all original $Imn22$
can be replaced by the maximum of lower-order $Imn22$ combinations
using the linear inequality method, which can be violated by some
entangled states in $Q_{a}$ and $Q_{b}$.

This work can help us understand the mathematical structures of Bell-CH
inequalities. A number of interesting topics are open for future studies.
One could investigate a set of Bell-CH inequalities for multipartite
systems using the method of the linear inequality. Appropriate ways
might be found to test our Bell-CH inequalities, especially inequality
(\ref{eq:nch}), in systems of optics, high-energy physics, or condensed
matter.
\begin{acknowledgments}
C.Q., Y.-G.Y., and Q.W. are supported in part by the National Natural
Science Foundation of China (NSFC) under Grants No. 11890713 (a subgrant
of Grant No. 11890710), No. 11947301, and No. 12047502. C.-F.Q. is
supported in part by the National Natural Science Foundation of China
(NSFC) under Grants No. 11975236 and No. 11635009.
\end{acknowledgments}

\bibliographystyle{apsrev4-1}
\bibliography{0D__arxiv-2101_02426v2_Bell-CH-manuscript-modified}

\onecolumngrid

\appendix

\section{Rearrangement inequality}

\label{sec:rearrange}The rearrangement inequality (or the permutation
inequality) is
\begin{equation}
x_{n}y_{1}+\cdots+x_{1}y_{n}\leqslant x_{\sigma(1)}y_{1}+\cdots+x_{\sigma(n)}y_{n}\leqslant x_{1}y_{1}+\cdots+x_{n}y_{n}
\end{equation}
for every permutation of $\sigma(i)$ ($i=1,\ldots,n$), with $n$
real numbers $x_{1},...,x_{n}$ which satisfy
\begin{align}
x_{1} & \leqslant\cdots\leqslant x_{n}
\end{align}
and $n$ real numbers $y_{1},...,y_{n}$ which satisfy
\begin{equation}
y_{1}\leqslant\cdots\leqslant y_{n}.
\end{equation}

If the numbers are different, that is to say, $x_{1}<\cdots<x_{n}$
and $y_{1}<\cdots<y_{n}$, then the lower bound is obtained only for
the permutation which reverses the order, i.e., $\sigma(i)=n-i+1$
for all $i=1,...,n$, and the upper bound is obtained only for the
identity permutation, i.e., $\sigma(i)=i$ for all $i=1,...,n$.

\section{Jensen's inequality}

\label{sec:jensen}Suppose $f(x)$ is a non-negative measurable function
satisfying
\begin{equation}
\int_{-\infty}^{\infty}f(x)dx=1,
\end{equation}
which is a probability density function in the probabilistic view.
Jensen's inequality about convex integrals is
\begin{equation}
\varphi\left(\int_{-\infty}^{\infty}g(x)f(x)dx\right)\leqslant\int_{-\infty}^{\infty}\varphi(g(x))f(x)dx
\end{equation}
for any real-valued measurable function $g(x)$ and ${\textstyle \varphi}$
is convex over $g(x)$. If $g(x)=x$, then Jensen's inequality reduces
to
\begin{equation}
\varphi\left(\int_{-\infty}^{\infty}xf(x)dx\right)\leqslant\int_{-\infty}^{\infty}\varphi(x)f(x)dx.
\end{equation}

\end{document}